\documentstyle[12pt,aps,prd,epsfig,preprint]{revtex}

\begin{document}
\draft
\def\beq{\begin{equation}}
\def\eeq{\end{equation}}
\def\bea{\begin{eqnarray}}
\def\eea{\end{eqnarray}}
\def\ba{\begin{array}}
\def\ea{\end{array}}
\def\nnb{\nonumber}
\def\ve{\vert}
\def\vel{\left|}
\def\ver{\right|}
\def\ds{\displaystyle}
\def\0{\c{s}}

\title{\bf Thermoelectric power of nondegenerate Kane
semiconductors under the conditions of mutual electron-phonon
drag in a high electric field}
\author{M. M. Babaev$^{a}$, T. M. Gassym$^{a,b}$, M. Ta\0$^{b}$   
\thanks{e-mail:~tasm@metu.edu.tr} and M. Tomak$^{b}$}
\address{$^a$Institute of Physics, Academy of Sciences of
Azerbaijan Baku 370143, Azerbaijan\\
$^b$Physics Department, Middle East Technical University 06531
Ankara, Turkey}

\maketitle
\begin{abstract}
The thermoelectric power of nondegenerate Kane semiconductors with 
due regard for the electron and phonon heating, and their thermal
and mutual drags is investigated. The electron spectrum is taken
in the Kane two-band form. It is shown that the nonparabolicity of
electron spectrum significantly influences the magnitude of the
thermoelectric power and leads to a change of its sign and
dependence on the heating electric field. The field dependence of
the thermoelectric power is determined analytically under various
drag conditions.
\end{abstract}
\thispagestyle{empty} 
~~\\ 
\narrowtext
\newpage
\setcounter{page}{1}

\section{Introduction}
Recently, the interest in thermoelectric power both theoretically
and experimentally in various systems, mesoscopic quantum
dots\cite{1,2}, quantum wires\cite{3}, heterojunctions and quantum
well structures\cite{4}-\cite{11} as well as the bulk
materials\cite{11,12}, has been intensified. Almost all of the
earlier theoretical investigations for analyzing the
diffusion\cite{3,15,16,17} and phonon drag\cite{7,8,9,18}
components of the thermoelectric power in macroscopic systems are
based on the Boltzmann equation. In these works, the weakly
nonuniform systems under the linear transport conditions are
considered in the absence of external electric field and in the
presence of lattice temperature gradient. 

There are some theoretical investigations of thermoelectric and 
thermomagnetic effects in semiconductors at high external electric
and nonquantizing magnetic fields\cite{19}-\cite{23}. In these
studies, heating of electrons and phonons, and their thermal and
mutual drags for the parabolic spectrum of nondegenerate electrons
and for the nonparabolic spectrum of degenerate electrons are
considered. These investigations are based on the solution of the
coupled system of kinetic equations of hot electrons and phonons in
nonlinear transport conditions. There are also theoretical
investigations of this problem in the hydrodynamic approximation.

Lei theoretically discussed the thermoelectric power of both bulk 
materials and quantum wells in the presence of charge carrier
heating with a high applied electric field by using the so-called
``balance equation approximation" for weakly nonuniform
systems\cite{11,13,24}. These calculations indicate that the hot
electron effect on the thermoelectric power may not only change
its magnitude but also change its sign at high electric fields.
This result has been confirmed by Xing {\it et al.}\cite{12} using
the nonequilibrium statistical operator method of Zubarev\cite{14}
jointly with the Lei-Ting balance equation approach\cite{24}. In
\cite{11} and \cite{12} the phonon drag contribution to
thermoelectric power is neglected at electron temperatures of
interest for hot electron transport. Thus, in both treatments this
contribution which is known to be important in linear transport at
low temperatures in bulk semiconductors\cite{10} and two-dimensional
systems\cite{4,5,6,10} is missed. By using the hydrodynamic balance
equation transport theory extended to weakly nonuniform systems, Wu
{\it et al}. carried out a calculation of the phonon drag
contribution to thermoelectric power of bulk semiconductors and
quantum well structures\cite{26}. According to the authors, the
balance equation approach has the advantage of easy inclusion of
hot electron effect and claims the importance of the phonon drag
contribution to thermoelectric power in hot electron transport
condition. They note that their consideration is applicable in the
regime where the electron drift velocity is lower than the sound
velocities for materials having high impurity concentrations and
intermediate electric field strength. Contrary to the assumptions
of Xing {\it et al}.\cite{12}, their results demonstrate that the
phonon drag contribution is remarkably enhanced at low lattice
temperature under the conditions considered. It is shown in
\cite{11} that the diffusion component of the thermoelectric power
may be negative within a low enough lattice temperature range at
high electric field while the phonon drag component is still
positive. In connection with these conclusions, it is necessary to
note that such a result was obtained in 1977 by Babaev and Gassymov
in \cite{20}. In that paper, the thermoelectric power and transverse
Nernst-Ettingshausen (NE) effect in semiconductors at high electric
and nonquantizing magnetic fields are studied by solving the coupled
system of kinetic equations for electrons and phonons. In the
investigation, both the heating of electrons and phonons, and the
phonon drag are taken into account. It is shown that when the
temperature gradient of hot electrons ($\nabla T_e$) is produced by
the lattice temperature gradient ($\nabla T$), $\nabla E=0$ and
${\ds \nabla T_e=(\partial T_e/\partial T)\nabla T}$, the electronic
parts of the thermoelectric and the NE fields reverse their sign.
In the case of heated phonons and $T_p=T_e \gg T$, both electronic
and phonon parts of the thermoelectric and thermomagnetic fields
reverse their sign for all cases considered. Here $T_e$, $T_p$ and
$T$ are the temperature of electrons, phonons and lattice,
respectively. In \cite{12} the thermoelectric power of charge 
carriers heated under a strong applied electric field in
semiconductors is obtained by making use of the nonequilibrium
statistical operator method. The final Eqs. (18) and (19) for
thermopower and the conclusion that the hot electron effect may
change both the magnitude and sign of the thermopower repeat the
results obtained in \cite{20} for a special case (when $\nabla T_e$
is realized by $\nabla T$). Moreover, we note that for the high
field case considered in \cite{12}, hot electrons (or semiconductor)
are in the regime of phonon generation. Therefore, both the
distribution function and the state of phonons are nonstationary as 
a result of the mutual drag of charge carriers and phonons at high
electric field, which is considered in \cite{27,28,29}. For the
role of the mutual electron-phonon drag and phonon generation at
high external electric and magnetic fields, see \cite{28,29,30}.

Recently, the interest in the study of thermoelectric and NE effect in 
II-VI semiconductors has been intensified\cite{31}-\cite{34}. Earlier 
investigations of the magnetic field dependence of the longitudinal NE
effect in HgSe\cite{35,36} and lead chalcogenides\cite{37,38} in the 
region of comparatively high temperatures ($T \geq 77 K$) demonstrated 
that the thermo emf exhibits saturation in the classical region of
strong magnetic fields $H$ irrespective of the dominant scattering
mechanism of charge carriers in the conduction band. However,
measurements of the longitudinal NE effect in iron-doped HgSe samples
at low temperatures ($20 \leq T \leq 60 K$), revealed presence of a
maxima in the change of thermoelectric power $\Delta\alpha(H)=\mid
\alpha(H)-\alpha(0) \mid$. $\Delta\alpha(H)$ first increases
quadratically with increasing $H$ for $\Omega\tau<1$, then passes
through a maximum for some $H=H_m$, and finally decreases as the
field increases further. Here, ${\ds \Omega=eH/(mc)}$ is the
cyclotron frequency, and $\tau$ is the electron relaxation time.
Another unusual fact is the sign reversal of the transverse NE
coefficient $Q_{\perp}(H)$ with magnetic field increasing in the
range $\Omega \tau>1$\cite{33,34}. The experiments in Ga-doped HgSe
demonstrated that at low temperatures, NE coefficients change sign
with increasing Ga~ concentration or the applied magnetic field
strength. The unusual features of the NE effect observed in HgSe
crystals may be attributed to the effect of mutual drag, which can
experimentally be detected in semiconductors with high concentration
of conduction electrons\cite{39}. As it is shown in the present
paper, these conditions can be realized more easily under high
external electric field at arbitrary temperatures.

A consistent microscopic theory of transport phenomena in
semiconductors and semimetals in high external electric and magnetic
fields with due regard for the heating of charge carriers and
phonons, their thermal and mutual drags, and the possible phonon
generation by the drift charge carriers must be based on the solution
of coupled system of kinetic equations for charge carriers and
phonons. Such a problem is formulated and solved for the first time
by Gassymov\cite{28}, see also reference \cite{27}. In the statement
of the problem, it should be noted that the traditional approximation
of small anisotropy of phonon distribution function (so-called
``diffusion approximation") is applicable to phonons whose drift
velocities ($u$) is much smaller than the sound velocity ($s_0$) in
crystal. In the presence of external electric and magnetic fields,
this condition obviously is not fulfilled. This violation shows up
particularly in several ways under the acoustical instability
conditions $(u \geq s_0)$. Actually, both spherically symmetric,
$N_s(q)$, and antisymmetric, $N_a(q)$, parts of the phonon
distribution function as well as ${\ds N_a(q)/N_s(q)}$ grow as $u$
increases. Indeed, ${\ds N_a(q)/N_s(q) \rightarrow 1}$
as $u \rightarrow s_0$, and ${\ds N_a(q)/N_s(q) \gg 1}$ when
$u \gg s_0$. The general solution of the Boltzmann equation for
phonons shows that $N(q)$ is stationary for $u<s_0$, and
nonstationary for $u \geq s_0$. These results are obtained by
solving the nonstationary kinetic equation for phonons interacting
with charge carriers at high electric and arbitrary magnetic
fields in the nondiffusion approximation\cite{27,28,29}.

In the light of the foregoing discussion, we must note that the
method of calculation used in \cite{11}, \cite{12} and \cite{26} has
intrinsically questionable assumptions. Actually in the process of
obtaining the force and energy balance equations, it is assumed that
the distribution function of electrons has the form of drifted Fermi
distribution function, and that of phonons has the form of drifted
Planck's distribution function with effective electron temperature
$T_e$ and electron drift velocity $v_d$ as a result of the
electron-phonon collisions. These assumptions mean that this method
is applicable only in the strong mutual drag conditions when
$\nu_p \gg \nu_i$ and $\beta_e \gg \beta_p$, i.e., electrons and
phonons transfer their energy and momentum to each other, and as a
result they have the same effective temperature and drift velocity.
Note that here $\nu_p$ and $\nu_i$ are the collision frequencies of
electrons with phonons and impurities, $\beta_e$ and $\beta_p$ are
the collision frequencies of phonons with electrons and phonons,
respectively. Under the strong mutual drag conditions, drift
velocities of electrons and phonons are the same, $u=s_0$, only at
the acoustical instability threshold (AIT). At AIT, the distribution
function of phonons is nonstationary and grows linearly in time. In
other words, drift velocities of electrons and phonons may be equal
to each other only at the nonstationary conditions of phonon
generation or amplification in external electric and magnetic
fields\cite{28,29}. Thus, the assumptions made in \cite{11},
\cite{12} and \cite{26} make it possible to use this method only
under the strong mutual drag conditions and in the region of drift
velocities $v_d \ll s_0$. On the other hand, under the mutual drag
conditions and $v_d \ll s_0$, electrons and phonons interacting
with electrons may have the same temperature $T_e=T_p$, but their
drift velocities may not be equal to each other, i.e., $v_d \neq u$. 

What about the terminology of thermal drag (or the drag of electrons 
by phonons), and mutual drag of electrons and phonons? There is a
misunderstanding. Actually, the terminology of mutual drag covers
the drag of electrons by phonons if $\nu_i \gg \nu_p$ and
$\beta_e \gg \beta_{pb}$ as well as the drag of phonons by electrons
if $\nu_p \gg \nu_i$ and $\beta_e \gg \beta_{pb}$. Here $\beta_{pb}$
is the collision frequency of phonons with phonons (p), and boundaries
of the crystal (b); and it is defined as $\beta_{pb}=\beta_p+\beta_b$.
Therefore, the mutual drag covers both the drag of electrons by
phonons (it is called ``thermal drag") and the drag of phonons by
electrons. The latter is named in the literature incorrectly as
``mutual drag". However, the mutual drag is the sum of both drags
and, for this reason, it is sometimes called as ``veritable drag".
In the mutual drag, electrons and phonons are scattered preferably
by each other, and the strong mutual drag may form a coupled
system with joint temperature $T_e=T_p$ and drift velocity $v_d=u$.

In the literature, usually the phonon drag effect (thermal drag) is
studied in the absence of heating external electric field and in
the presence of small $\nabla T$ in impure semiconductors when the
collision frequency of electrons with impurity ions is much greater
than that of electrons with phonons (low mobility, low temperature
and high impurity concentration). In this situation the drag of
phonons by electrons is less than the drag of electrons by phonons
(thermal drag). In high external electric field, electrons are
heated and the frequency of their scattering by impurity ions
decreases; meanwhile their scattering frequency by phonons increases. 

For the nondegenerate hot electrons with parabolic spectrum and
effective temperature $T_e$, the ratio ${\ds \nu_i/\nu_p \sim
(T_e/T)^{-3}}$ decreases sharply, and becomes unity at some critical
value of the electric field $E=E_{cr}$. For $E>E_{cr}$, electrons
and phonons scatter from each other, and the effect of their mutual
drag becomes important. The experiments for investigation of the
effect of phonon drag in specimens of InSb or Ge are usually carried
out at external fields $E>10~Vcm^{-1}$ and lattice temperatures
$T<20~K$. At these conditions $T_e \approx 10^2,10^3~ T$. 

The effect of high electric field is not limited by the heating of 
electrons; it also leads to the following effects: \\
{\bf a.}~~The drift velocity of electrons increases. Indeed, when 
$\nabla T_e \parallel \nabla T$, $v_d \gg {\ds v_{\nabla T}}$.
Here ${\ds v_{\nabla T}}$ is the drift velocity of phonons in the
presence of $\nabla T$.\\
{\bf b.}~~The ratio ${\ds \beta_e/\beta_p}$ increases as 
${\ds T_e/T}$ increases.\\
{\bf c.}~~The momentum range of phonons interacting with
electrons increases by $T_e$ as $0<q<2 {\ds \bar{p}=
\sqrt{8m T_e} \equiv 2p_T (T_e/T)^{1/2}}$.\\
{\bf d.}~~The number of phonons interacting with electrons
increases by $T_e$ linearly. Namely, ${\ds N(q)=T_e/
(\hbar\omega_q^{\star})}$. This is the most important finding.\\
{\bf e.}~~Under the mutual drag conditions, the inelasticity of
scattering of electrons by phonons is obtained from
${\ds \hbar \omega_q^{\star}=\hbar\omega_q-{\bf u}{\bf q}}$.
It decreases with increasing $u$, and ${\ds N(q)=N(q,T_e)/
(1-{\bf u}.{\bf q}/\hbar\omega_q)}$ increases as $u$
increases. Because,
the denominator goes to zero as $u\rightarrow s_0$. At these
drift velocities, the phonon generation or amplification by the
external electric field starts, and the state of phonons becomes
nonstationary. Under these conditions the thermal drag, which is
proportional to the degree of the inelasticity of the
electron-phonon scattering, tends to zero, and the mutual drag
of electrons and phonons is strong. Therefore, electrons and
phonons form a system coupled by the mutual drag with common
temperature $T_e$ and drift velocity ${\bf u}$\cite{27,28,29}.

The organization of the paper is as follows. The theoretical
analysis of the problem is given in Sec. II. In Sec. III we
discuss the results of the present work in detail. Finally, the
conclusion is given in Sec. IV.

\section{Theory}
Two-band Kane spectrum of electrons is:
\beq
p(\varepsilon)=(2m_n\varepsilon)^{(1/2)}
\left(1+\frac{\varepsilon}{\varepsilon_g}\right)^{(1/2)},
\eeq
where $m_n$ is the effective mass of electrons at the bottom of
the conduction band, $\varepsilon_g$ is the band gap, $p$ and
$\varepsilon$ are the electron momentum and energy,
respectively\cite{17}.  

The physical process considered is the thermoelectric Seebeck
effect in the presence of a heating electric field ${\bf E}$ and
$\nabla T_e$, which can be produced by $\nabla E$ or $\nabla T$.

The basic equations of the problem are the coupled Boltzmann
transport equations for electrons and phonons. The quasi-elastic
scattering of electrons by acoustic phonons is considered. For the
case considered, the distribution functions of electrons
$f({\bf p},{\bf r})$ and phonons $N({\bf q},{\bf r})$ may be
presented in the form:
\beq
f({\bf p},{\bf r})=f_0(\varepsilon,{\bf r})+{\bf f_1}(\varepsilon,
{\bf r}){\frac{{\bf p}}{p}},~~~~~~~\left|{\bf f_1}\right| \ll f_0,
\eeq
\beq
N({\bf q},{\bf r})=N_0(q,{\bf r})+{\bf N_1}(q,{\bf r})
{\frac{{\bf q}}{q}},~~~~~~~\left|{\bf N_1}\right| \ll N_0.
\eeq
Here $f_0$ and ${\bf f_1}$, $N_0$ and ${\bf N_1}$ are the isotropic 
and the anisotropic parts of the electron and phonon distribution 
functions, respectively.

If the inter-electronic collision frequency $\nu_{ee}$ is much
greater than the collision frequency of electrons for the energy
transfer to lattice $\nu_{\varepsilon}$, then
$f_0(\varepsilon,{\bf r})$ is the Fermi distribution function
with an effective electron temperature $T_e$. We consider the
case that there is a ``thermal reservoir" of short-wavelength
(SW) phonons for the long-wavelength (LW) phonons, with maximum
quasi-momentum ${\ds q_{max} \approx 2 \overline{p} \ll T/s_0}$,
interacting with electrons. In this case $N_0(q,{\bf r})$ has
the form:
\beq
N_0(q,{\bf r})\approx \frac{T_p({\bf r})}{s_0q},
\eeq
where $T_p$ is the effective temperature of LW phonons\cite{40}. 

Starting from the Boltzmann transport equations, we obtain the
following relations for ${\bf f_1}$ and ${\bf N_1}$ in the steady
state:
\beq
\frac{p}{m(\varepsilon)} \nabla f_0 - e {\bf E_c}~
\frac{p}{m(\varepsilon)} \frac{\partial f_0}{\partial \varepsilon}
+\nu(\varepsilon){\bf f_1}+
\frac{2 \pi m(\varepsilon)}{(2 \pi \hbar)^3~p^2}
\frac{\partial f_0}{\partial \varepsilon}
\int_{0}^{2p} {\bf N_1}(q) W(q) \hbar \omega_q q^2 ~dq =0,
\eeq
\beq
S_0 \nabla N_0 + \beta(q) {\bf N_1}- \frac{4 \pi m(\varepsilon)}
{(2 \pi \hbar)^3} W(q)N_0(q)\int_{q/2}^{\infty} {\bf f_1}~dp =0,
\eeq
where $e$ is the absolute value of the electronic charge, ${\bf E_c}
={\bf E}+{\bf E_T}$, with ${\bf E_T}$ as thermoelectric field,
$m(\varepsilon)$ is the effective mass of electron,
$\hbar\omega_q=s_0 q$ is the phonon energy, ${\ds W(q)=W_0 q^t}$ is
the square of the matrix element of the electron-phonon interaction
($t=1$ for deformation and $t=-1$ for piezoelectric interaction),
$\beta(q)$ and $\nu(\varepsilon)$ are the total phonon and electron
momentum scattering rates, respectively.
 
For the Kane semiconductors with electron spectrum given by Eq. (1), 
$m(\varepsilon)$ and $\nu(\varepsilon)$ have the form\cite{17}:
\beq
m(\varepsilon)=m_n \left(1+\frac{2 \varepsilon}{\varepsilon_g}\right),
\eeq
\beq            
\nu(\varepsilon)=\nu_0(T)\left(\frac{T_p}{T}\right)^l \left(1+
\frac{2\varepsilon}{\varepsilon_g}\right)\left(1+\frac{\varepsilon}
{\varepsilon_g}\right)^{-r}\left(\frac{\varepsilon}{T}\right)^{-r},
\eeq
where $r=3/2$, $l=0$ for the scattering of electrons by impurity
ions, and $r=-t/2$, $l=1$ for the scattering of electrons by acoustic
phonons. When LW phonons are scattered by SW phonons or by crystal
boundaries, $\beta(q)$ does not depend on the spectrum of electrons
and has the form\cite{40}:
\beq
\beta_p(q)=\frac{T^4}{4 \pi \rho \hbar^4 s_0^4}~ q,~~~~~ 
\beta_b(q)=\frac{s_0}{L},
\eeq
where the indices $p$ and $b$ denote the scattering of LW phonons
by SW phonons and crystal boundaries, $\rho$ and $L$ are the
density and the minimum size of specimen, respectively. On the
other hand, when LW phonons are scattered by electrons, $\beta_e(q)$
depends on the spectrum of electrons, and for the spectrum given by
Eq. (1) we obtain:
\beq
\beta_e(q)= \left(\frac{m_n s_0^2}{8 \pi T_e} \right)^{1/2} 
\frac{N W_0}{T_e} \left(1+\frac{2T_e}{\varepsilon_g} \right)^2
\left(1+\frac{3T_e}{2 \varepsilon_g} \right)^{-3/2}q^t,
\eeq
where $N$ is the concentration of electrons.

Solving the coupled Eqs. (5) and (6) by the same way as in
\cite{23}, it is easy to calculate the electric current density
of electrons\cite{17},
\beq
{\bf J}=-\frac{e}{3 \pi^2
\hbar^3}\int_{0}^{\infty}{\bf f_1}(\varepsilon) p^2
(\varepsilon)~d\varepsilon.
\eeq

Let the external electric field be directed along the $x$ axis,
and $\nabla T$ (or the external electric field gradient
$\nabla E$) along the $z$ axis. Under these conditions the
electron part $(\alpha_e)$ and phonon part $(\alpha_p)$ of the
thermoelectric power $(\alpha)$ are obtained from equation
$J_z=0$ as:
\beq
\alpha=\alpha_e+\alpha_p~;~~~~~ \alpha_e=-\frac{\beta_{11}^{(e)}}
{\sigma_{11}}~; ~~~~~\alpha_p=-\frac{\beta_{11}^{(p)}}{\sigma_{11}},
\eeq
where
\beq
\sigma_{11}=\int_{0}^{\infty}a(x)[1+b(x)]~dx,
\eeq
\beq
\beta_{11}^{(e)}=\frac{1}{e}\int_{0}^{\infty} a(x)
\left\{x-\frac{\zeta(T_e)}{T_e}+\left[1-\frac{\zeta(T_e)}{T_e}
\right]b(x)\right\}dx,
\eeq
\beq
\beta_{11}^{(p)}=\frac{1}{e}\int_{0}^{\infty} a(x)
\left\{\lambda(x)+\lambda(\vartheta_e) b(x)\right\}dx,~~~
x=\frac{\varepsilon}{T_e},~~\vartheta_e =\frac{T_e}{T},~~
\vartheta_p = \frac{T_p}{T}. 
\eeq
here $\zeta(T_e)$ is the chemical potential of hot electrons,
\beq
a(x)=\frac{e^2}{3\pi^2\hbar^3}~ \frac{p^3(x)}{m(x)\nu(x)}
~\exp\left[\frac{\zeta(T_e)}{T_e}-x\right],
\eeq
\beq
b(x)= \frac{\gamma(x)}{1-\gamma(\vartheta_e)}~
\frac{m(x)}{m(\vartheta_e)}~\frac{\nu(x)}{\nu(\vartheta_e)},
\eeq
\beq
\gamma(x)=\frac{3+t}{(2p)^{3+t}}~\frac{\nu_p(x)}{\nu(x)}
\int_{0}^{2p}\frac{\beta_e(q)}{\beta(q)}~q^{2+t}~dq,
\eeq
\beq
\lambda(x)=\frac{3+t}{(2p)^{3+t}}~\frac{m(x)s_0^2}{T_p}
~\nu_p(x) \int_{0}^{2p}~\frac{1}{\beta(q)}~q^{2+t}~dq,
\eeq
where $\nu_p(x)$ is the scattering frequency of electrons by
phonons. The coefficient $\lambda(x)$ characterizes the efficiency
of the thermal drag, and $\gamma(x)$ describes the same for the
mutual drag.

As it follows from Eq. (12), by taking into account Eqs. (13)-(15), 
$\alpha_p$ consists of ``thermal drag" and ``mutual drag" terms. 
Actually, the first term in Eq. (15) considers ``the drag of
electrons by phonons" (thermal drag) and the second term considers
``the drag of phonons by electrons" (mutual drag).

In Eq. (15), the first term is dominant if~ $\nu_i \gg \nu_p$~ and 
$\beta_e \gg \beta_{pb}$, i.e., phonons are scattered preferably
by electrons, but electrons are scattered by impurity ions
(thermal drag). The second term is dominant, on the other hand,
if~ $\nu_i \ll \nu_p$ and $\beta_e \gg \beta_{pb}$. Since at high
electric fields~ ${\ds \nu_i(\varepsilon)/\nu_p(\varepsilon)=
\nu_i(T)/\nu_p(T)\left(T_e/T\right)^{-3}=E_{cr}/E}$,~ the mutual
drag dominates for~ $E>E_{cr}$. Using the total collision frequency
$\nu(\varepsilon)=\nu_i(\varepsilon)+\nu_p(\varepsilon)$, we study
$E$ dependence of the thermal and mutual drags by using Eq. (15).

The ratio of the second and first terms in Eq. (15) is
${\ds\left[\lambda(\vartheta)/\lambda(x)\right]b(x)}$. When
${\ds x=\bar{x}=T_e/T}$, $\left[\lambda(\vartheta)/\lambda(\bar{x})
\right]=1$. Therefore, we have ${\ds \left[\lambda(\vartheta)/
\lambda(\bar{x})\right]b(\bar{x})\approx b(\vartheta)=
\gamma(\vartheta)/\left[(1-\gamma(\vartheta)\right]}$. As it
follows from this result, ${\ds \gamma(\vartheta)/
\left[1-\gamma(\vartheta)\right]}$ is smaller than 1 for
$1/2<\gamma(\vartheta)<1$, equal to 1 for $\gamma(\vartheta)=1/2$,
and larger than 1 for $1/2<\gamma(\vartheta)<1$. Moreover, it tends
to infinity as $\gamma(\vartheta) \rightarrow 1$. Therefore, at
high electric field the mutual drag is more important.

Because of the complexity of general analysis of Eqs. (12)-(15), 
hereafter we examine the dependence of electron momentum on its 
energy in the form:
\beq
p(\varepsilon)=(2m_n \varepsilon_g)^{1/2} 
\left(\frac{\varepsilon}{\varepsilon_g}\right)^{s}.          
\eeq
This form, for the spectrum given by Eq. (1), corresponds to
parabolic case for $T_e \ll \varepsilon_g$, $s=1/2$, and
strongly nonparabolic case for $T_e \gg \varepsilon_g$, $s=1$.
In these cases $m(\varepsilon)$, $\nu(\varepsilon)$ and
$\beta(q)$ may be presented as:
\beq
m(\varepsilon)=2sm_n\left(\frac{\varepsilon}
{\varepsilon_g}\right)^{2s-1},
\eeq
\beq
\nu(\varepsilon)=2s \nu_0(T) \vartheta_p^l\left(\frac{\varepsilon}
{\varepsilon_g}\right)^{(2s-1)(1-r)}\left(\frac{\varepsilon}{T}
\right)^{-r},
\eeq
\beq
\beta(q)=\beta(T) \vartheta_e^{n(s-2)} \left(\frac{T}{\varepsilon_g}
\right)^{n\left(s-1/2\right)}\left(\frac{s_0 q}{T}\right)^k,
\eeq
where $n=1$, $k=t$ for scattering of LW phonons by electrons, $n=0$, 
$k=0$ for scattering by the crystal boundaries, and $n=0$, $k=1$ for 
scattering by SW phonons.

For the spectrum expressed by Eq. (20), from Eqs. (12)-(19) we obtain:
\beq
\alpha_e=-\frac{1}{e}\left(1+C_1 \frac{\gamma_0}
{1-\gamma_0}\right)^{-1}\left\{3-s+2sr-\frac{\zeta(T_e)}{T_e}+
\left[1-\frac{\zeta(T_e)}{T_e}\right]
C_1\frac{\gamma_0}{1-\gamma_0}\right\},
\eeq	
\bea
\alpha_p =-\frac{1}{e} ~\frac{C_2+(C_1-C_2)\gamma_0}
{1+(C_1-1)\gamma_0}~\frac{(3+t)~2^{(2-\frac{3k}{2})}s^2}
{3+t-k}\left(\frac{m_ns_0^2}{T}\right)^{(1-k/2)}\\
\nonumber \left(\frac{T \vartheta_e}{\varepsilon_g}
\right)^{(s-1/2)(4+t-k-n)}\vartheta_e^{(3n+t-k)/2)}
~\frac{\nu_{p0}(T)}{\beta(T)},
\eea
where
\beq
C_1=\frac{\Gamma(1+3s+2sr+2st-sk)}{\Gamma(3-s+2sr)},~~~~
C_2=\frac{\Gamma(1+3s+2sr+st-sk)}{\Gamma(3-s+2sr)},
\eeq
\bea
\gamma_0 = \frac{(3+t) 2^{\frac{3 (t-k)}{2}}}{3+2t-k}
\left(\frac{m_n s_0^2}{T} \right)^{(\frac{t-k}{2})} 
\left(\frac{T \vartheta_e}{\varepsilon_g}
\right)^{(s-1/2)(2r+2t-k-n+1)}\\ 
\nonumber \vartheta_e^{(r+t+(3n-3-k)/2)} 
~\vartheta_p^{1-l}~\frac{\beta_e(T)}{\beta(T)} 
~\frac{\nu_{p0}(T)}{\nu_0(T)}.
\eea

The chemical potential of nondegenerate electrons for the
spectrum in Eq. (20) becomes:
\beq
\zeta(T_e)=T_e~\ln\left\{\frac{3\pi^2\hbar^3N}{\Gamma(1+3s)
(2m_nT)^{3/2}}\left(\frac{T}{\varepsilon_g}
\right)^{-3\left(s-1/2\right)}\vartheta_e^{-3s}\right\}.
\eeq

Consider the limits $\gamma_0 \ll 1$ and $\gamma_0 \rightarrow 1$.
The first limit corresponds to the weak mutual drag case. In this
case, by using Eqs. (24) and (25), the components of the
thermoelectric power is found to be:
\beq
\alpha_e=-\frac{1}{e}\left\{3-s+2sr-\frac{\zeta(T_e)}{T_e}
-C_1(2-s+2sr) \gamma_0\right\},
\eeq
and
\bea
\alpha_p =-\frac{1}{e}~\left\{C_2+C_1(1-C_2)\gamma_0\right\}
~\frac{(3+t)2^{(2-\frac{3k}{2})}s^2}{3+t-k}
\left(\frac{m_ns_0^2}{T}\right)^{(1-k/2)}\\
\nonumber \left(\frac{T \vartheta_e}{\varepsilon_g}
\right)^{(s-1/2)(4+t-k-n)}\vartheta_e^{(3n+t-k)/2)}
~\frac{\nu_{p0}(T)}{\beta(T)}.
\eea

Since $C_1>0$, and $2-s+2sr \geq 0$ for all real scattering
mechanisms and the spectrum of electrons with $s \geq 1/2$,
from Eq. (29) we find that the mutual drag leads to a decrease
of $\alpha_e$ both in the parabolic and nonparabolic cases.

The $\gamma_0 \rightarrow 1$ limit, on the other hand,
corresponds to the strong mutual electron-phonon drag. In this
case $k=t$, $n=1$, $r=-t/2$, $l=1$, and $\vartheta_p=\vartheta_e$.
From Eq. (27) we obtain ${\ds \gamma_0=[\beta_e(T)/\beta(T)]
[\nu_{p0}(T)/\nu_0(T)]\rightarrow 1}$. Hence, $\alpha_e$ and
$\alpha_p$ take the form:
\beq
\alpha_e=-\frac{1}{e} \left\{1-\frac{\zeta(T_e)}{T_e}\right\},
\eeq 
\beq
\alpha_p=-\frac{1}{e} \frac{4\sqrt{2}~(2s)^2}{3\pi^{3/2}} 
\left(\frac{T}{\varepsilon_g}\right)^{3 \left(s-1/2\right)}
~\frac{(m_nT)^{3/2}}{\hbar^3 N}~\vartheta_e^{3s}.
\eeq

One can also see the decrease of $\alpha_e$ by the influence of
mutual drag, from a comparison of Eqs. (31) and (29). As it
follows from Eq. (28), for nondegenerate electrons we have:
\beq
\frac{(m_nT)^{3/2}}{\hbar^3 N}\left(\frac{T}{\varepsilon_g}
\right)^{3\left(s-1/2\right)} \approx 
~\exp \left[-\frac{\zeta(T)}{T} \right] \gg 1.
\eeq

The $E$ dependence of $\vartheta_e$ in the weak mutual drag case
was considered elsewhere\cite{21}. Here we investigate the same
dependence in the strong mutual drag conditions. In this case the
electron temperature is determined by the energy balance equation:
\beq
\sigma_{11}(\vartheta_e) E^2=W_{pp}(\vartheta_e),
\eeq
where $W_{pp}(\vartheta_e)$ is the power transferred by LW
phonons to the ``thermal reservoir" of SW phonons. Now we consider
the following limiting cases:
\beq
i.~~~\frac{\beta_p+\beta_b}{\beta_e} \ll \frac{\nu_i}{\nu_p},~~~~~~~
ii.~~~\beta_p \gg \beta_b,~~~
\frac{\beta_p}{\beta_e} \gg \frac{\nu_i}{\nu_p},~~~~~~~
iii.~~~\beta_p \ll \beta_b,~~~
\frac{\beta_b}{\beta_e} \gg \frac{\nu_i}{\nu_p}. 
\eeq
The results obtained for $\vartheta_p=\vartheta_e \gg 1$ are given
in Table I. 

As it is seen in Table I, the nonparabolicity of the electron
spectrum strongly changes $E$ dependence of the electron
temperature. Using Table I, one can easily obtain $E$ dependence
of $\alpha$ for the cases considered in Eq. (35). For instance, if
the first inequality is satisfied, then ${\ds\alpha_p \sim E^2}$
for the parabolic, and ${\ds \alpha_p\sim E^{3/2}}$ for the strong
nonparabolic spectrum of electrons.

Let us consider the dependences of $V_e$, $\alpha_p$ and $V_p$ on
$E$ for different scattering mechanisms of electrons and phonons.
As it follows from the results obtained above, the dependence of
$\alpha_e$ on $\vartheta_e$ or $E$ is weak (logarithmic) for the
limiting cases $\gamma_0 \rightarrow 0$ and
$\gamma_0 \rightarrow 1$. If $\vartheta_e \gg 1$ at one end of the
specimen, and $\vartheta_e=1$ at the other end, $V_e \sim
\vartheta_e$ by the accuracy of logarithmic dependence. When
$\gamma_0 \rightarrow 1$, ${\ds \alpha_p \sim \vartheta_e^{3s}}$
and ${\ds V_p \sim \vartheta_e^{3s+1}}$. 

Taking into account the foregoing discussion and Table I, one can
find the dependences of $V_e$, $\alpha_p$ and $V_p$ on $E$ as
$\gamma_0 \rightarrow 1$. The results are given in Table II.

In the weak mutual drag case, for $T_p=T_e \gg 1$, $\alpha_p$ and 
$\vartheta_e$ are given by:
\beq 
\alpha_p \sim \vartheta_e^{(4+t-k-n)+2n-2},~~~~~
\vartheta_e=\left(\frac{E}{E_i}\right)^{2/(8s-1-2rs+\ell)}, 
\eeq
where $E_i$ is: 
\beq
E_i=\left(\frac{T}{\varepsilon_g}\right)^{(s-1/2)(4-r)}
\left(\frac{m_n T}{\hbar^2 N^{2/3}}\right)^{3/4}\left(\frac{m_n T}
{e^2}\right)^{1/2}\left[\nu_e(T)\beta_p(T) \right]^{1/2}.
\eeq

We find dependence of $V_e$ on $E$ for several interaction
mechanisms as shown in Table III. 

In the weak mutual drag case, we obtain the $E$ dependence of
$\alpha_p$ and $V_p$ for several scattering mechanisms as follows:\\
{\bf 1.} Electrons are scattered by deformation acoustical (DA)
phonons; phonons transfer their energy to electrons, but momentum 
to the crystal boundaries. $t=1$, $r=-1/2$, $\ell=1$, $k=1$, 
$n=1$ (drag of phonons by electrons case):
\bea
\alpha_p ~~\sim ~E^{2/9}~~(s=1/2), ~~~~\sim~E^{2/3}~~(s=1),\\
\nonumber V_p ~~\sim ~E^{2/3}~~(s=1/2), ~~~~\sim ~E^{8/9}~~(s=1).
\eea

{\bf 2.} Electrons are scattered by DA phonons, and phonons by 
electrons. $t=1$, $r=-1/2$, $\ell=1$, $k=1$, $n=1$ (the mutual 
drag case):
\bea
\alpha_p ~~\sim ~E^{2/3}~~(s=1/2), ~~~~\sim~E^{2/3}~~(s=1),\\
\nonumber V_p ~~\sim ~E^{10/9}~~(s=1/2), ~~~~\sim~E^{8/9}~~(s=1). 
\eea

{\bf 3.} Electrons are scattered by piezo acoustical (PA) phonons; 
phonons transfer their energy to electrons and momentum to the 
crystal boundaries. $t=-1$, $r=1/2$, $\ell=1$, $k=0$, $n=0$ (drag 
of phonons by electrons case): 
\bea
\alpha_p ~~\sim ~E^{-2/7}~~(s=1/2), ~~~~\sim~E^{2/7}~~(s=1),\\
\nonumber V_p ~~\sim ~E^{2/7}~~(s=1/2), ~~~~\sim~E^{4/7}~~(s=1).
\eea

{\bf 4.} Electrons are scattered by PA phonons, and phonons by 
electrons. $t=-1$, $r=1/2$, $\ell=1$, $k=-1$, $n=1$ (the mutual 
drag case):
\bea
\alpha_p ~~\sim ~E^{6/7}~~(s=1/2), ~~~~\sim~E^{6/7}~~(s=1),\\
\nonumber V_p ~~\sim ~E^{10/7}~~(s=1/2), ~~~~\sim~E^{8/7}~~(s=1).
\eea

{\bf 5.} Electrons transfer their momentum to impurity ions,
energy to DA phonons; and phonons transfer their energy to
electrons, momentum to the boundaries. $t=1$, $r=3/2$, $\ell=0$,
$k=0$, $n=0$ (``thermal drag", or, drag of electrons by phonons):
\bea
\alpha_p ~~\sim ~E^{2/3}~~(s=1/2), ~~~~\sim~E^{3/2}~~(s=1),\\
\nonumber V_p ~~\sim ~E^2~~(s=1/2), ~~~~\sim~E^2~~(s=1).
\eea

{\bf 6.} The momentum of electrons is transferred to impurity
ions, energy to DA phonons; and phonons transfer their energy
and momentum to electrons. $t=1$, $r=3/2$, $\ell=0$, $k=1$,
$n=1$ (drag of electrons by phonons, or, ``thermal drag" case):
\bea
\alpha_p ~~\sim ~E^2~~(s=1/2), ~~~~\sim~E^{3/2}~~(s=1),\\
\nonumber V_p ~~\sim ~E^{10/3}~~(s=1/2), ~~~~\sim~E^2~~(s=1).
\eea

{\bf 7.} The momentum of electrons is transferred to impurity
ions, energy to PA phonons; and phonons transfer their energy
to electrons and momentum to the boundaries. $t=-1$, $r=3/2$,
$\ell=0$, $k=0$, $n=0$ (drag of electrons by phonons
``thermal drag"): 
\bea
\alpha_p ~~\sim ~E^{-2/3}~~(s=1/2), ~~~~\sim ~E^{1/2}~~(s=1),\\
\nonumber V_p ~~\sim ~E^{2/3}~~(s=1/2), ~~~~\sim ~E~~(s=1).
\eea

{\bf 8.} The momentum of electrons is transferred to impurity
ions, energy to PA phonons; and phonons transfer their energy
and momentum to electrons. $t=-1$, $r=3/2$, $\ell=0$, $k=-1$,
$n=1$ (``thermal drag" case):
\bea
\alpha_p ~~\sim ~E^2~~(s=1/2),~~~~\sim~E^{3/2}~~(s=1),\\
\nonumber V_p ~~\sim ~E^{10/3}~~(s=1/2),~~~~\sim~E^2~~(s=1).
\eea

It should be noted that the cases 6 and 8 lead to the same
results, because in both cases $r=3/2$, $\ell=1$, $k=t$, and
$n=1$.

\section{Discussion}
The nonparabolicity of electron spectrum significantly influences
the thermoelectric power of hot charge carriers and leads to a
change of its electron temperature dependence, as it is seen from
Eqs. (24) and (25). For all scattering mechanisms $4+t-k-n>0$.
Therefore, the nonparabolicity of the spectrum leads to a more
rapid increase of $\alpha_p$ with increasing $T_e$. Moreover,
$\alpha_p$ consists of the factor~
${\ds \nu_{p0}(T)/\beta(T) \gg 1}$.

As it follows from Eqs. (29) and (30), in the weak mutual drag
case $\alpha_e$ does not depend on $T_e$ or $E$ by the accuracy
of logarithmic dependence, and the thermoelectric field (or
voltage) depends on $T_e$ linearly. Indeed,
$\alpha_e \ll \alpha_p$, and $\alpha_p$ depends on $T_e$ and
$E$ more strongly. 

For nondegenerate electrons, the factor in Eq. (31) is:
\beq
\frac{(m_n T)^{3/2}}{\hbar^3 N}\left(\frac{T}
{\varepsilon_g}\right)^{3(s-1/2)}~~ \approx 
~~\exp\left(-\frac{\zeta(T)}{T}\right) \gg 1.
\eeq
By comparing Eqs. (31) and (32) we may easily see that under the
strong mutual drag condition, $\alpha_e \ll \alpha_p$. In other
words, the thermoelectric power mainly consists of the phonon
part. Indeed, we again see that the nonparabolicity of the
electron spectrum strongly changes the dependence of $\alpha_p$
on $T_e$. In the weak mutual drag case,
$\alpha_p \sim T_e^{(3n+t-k)/2}$ for the parabolic,and
$\alpha_p \sim T_e^{(2+n-k-t)}$ for the strong nonparabolic
spectrum of electrons. In the strong mutual drag,
$\alpha_p \sim T_e^{3/2}$ for the parabolic, and
$\alpha_p \sim T_e^3$ for the strong nonparabolic spectrum cases. 

According to Eq. (31) in the strong mutual drag case, the
dependences of $\alpha_e$ on $\vartheta_e$ and $E$ are logarithmic
and $V_e \sim \vartheta_e$. In Table I we see that under the strong
mutual drag conditions, $V_e$, $\alpha_p$ and $V_p$ grow as $E$
increases in the limiting cases given in Eq. (35). According to
Table II in the strong mutual drag case, the nonparabolicity of the
spectrum leads to a weaker dependence of $V_e$ on $E$ than in the
parabolic one. In other words, as $E$ increases, $V_e$ grows faster
in the parabolic case. The influence of the nonparabolicity of the
spectrum on $\alpha_p$ and $V_p$ is more complicated. In the Case i,
$\alpha_p$ and $V_p$ grow more rapidly with $E$ for the parabolic
spectrum. However, in the Case ii and Case iii, $\alpha_p$ grows
more rapidly with $E$ for the nonparabolic spectrum. On the other
hand, the dependence of $V_p$ on $E$ approximately is the same for
both parabolic and nonparabolic spectrum of electrons.

In the weak mutual drag case, According to Table III, for the
scattering of electrons by phonons, if $V_e$ is proportional to
$E^n$ for the parabolic spectrum, then, it is proportional to
$E^{2n}$ for the nonparabolic spectrum of electrons.

What about the dependences of $\alpha_p$ and $V_p$ on $E$ for the
weak mutual drag case? One can see from Eqs. (38)-(45) that for all
the cases considered, the thermoelectric voltage $V_p$ grows as $E$
increases. 

The cases 2 and 4 consider the mutual drag condition for the region
of common drift velocities $u \ll s_0$. In this case the dependence
of $\alpha_p$ on $E$ is exactly the same for both parabolic and
nonparabolic spectrums. But, the dependences of $V_p$ are different.
Actually, $V_p$ increases faster for the parabolic spectrum with
increasing $E$.

The cases 1 and 3 consider the drag of phonons by electrons under
the conditions of scattering of electrons by DA and PA phonons. As
it is seen from Eqs. (38) and (40), in these cases $\alpha_p$ and
$V_p$ grow more rapidly as $E$ increases for the nonparabolic
spectrum.

The cases 6 and 8 consider the drag of electrons by phonons or the 
``thermal drag". As it follows from Eqs. (43) and (45), the 
dependences of $\alpha_p$ and $V_p$ on $E$ are the same independent 
of the type of the scattering of electrons by DA or PA phonons. 
Moreover, $\alpha_p$ and $V_p$ grow faster as $E$ increases for the 
parabolic spectrum.

In cases 5 and 7 we have the condition of drag of electrons by
phonons with common drift velocities equal to that of phonons $u$.
In the case 5, the dependence of $V_p$ on $E$ is the same for both
the parabolic and nonparabolic spectrums, whereas $\alpha_p$ grows
more rapidly for nonparabolic case. On the other hand, both
$\alpha_p$ and $V_p$ grow faster for the nonparabolic spectrum as
$E$ increases in the case 7.

In the weak mutual drag case, $\vartheta_e$ is proportional to 
$E^{s[4+(t-k)-n]+2n-2}$. Therefore, when $t=k$ and $n=1$ we have 
${\ds \vartheta_e \sim E^{3s}}$.

In the absence of mutual drag, electronic part of the
thermoelectric field (or the integral thermoelectric power) is:
\beq
E_{cz}=-\frac{1}{e}\left(2rs-4s+3 \right)\nabla_z T_e.
\eeq
For the strong nonparabolic spectrum, when electrons are scattered
by PA phonons ($r=1/2$), $E_{cz}$ vanishes. However, when electrons
are scattered by DA phonons ($r=-1/2$), $E_{cz}$ reverses its sign
compared to the parabolic spectrum case. Thus, the nonparabolicity
of the electron spectrum leads to a change of the sign of the
thermoelectric field. 

In the case of the parabolic spectrum and heated electrons, if the 
electron temperature gradient is produced by the lattice
temperature gradient, then the electronic part of the 
thermoelectric field reverses its sign in comparison to the case
of nonheated electrons ($T_e=T$). For the case $T_p=T_e \gg T$,
$(\partial T_e/\partial T)<0$ is negative. Therefore,
both electronic and phonon parts of the thermoelectric field
reverse their signs compared to the nonheating case for all
considered situations.

\section{Conclusion}
In the present work, we show that the nonparabolicity of electron 
spectrum significantly influences the magnitude of the
thermoelectric power and leads to a change of its sign compared
to the parabolic spectrum case. The nonparabolicity also
remarkably changes the heating electric field dependence of the
thermoelectric power.

It is shown that in the strong mutual drag conditions, the electron
part of the thermoelectric power dominates over the phonon part.
Indeed, the thermoelectric power increases with the electronic
temperature as ${\ds \sim T_e^{3/2}}$ for the parabolic, and as
${\ds \sim T_e^3}$ for the strong nonparabolic spectrum of electrons.
For all the cases considered $\alpha_p$, and the thermoelectric
fields $V_e$ and $V_p$ grow as $E$ increases. Indeed, we show that
this grow is more rapidly for the parabolic spectrum of electrons.

In the weak mutual drag case for the scattering of electrons by
phonons, it is found out that $V_e$ grows faster with increasing
$E$ for the parabolic spectrum case. Moreover, for all the cases
studied $V_p$ grows as $E$ increases.

It is shown that in both weak and strong mutual drag cases,
electronic part of the thermoelectric power does not depend on
$T_e$ or $E$ by the accuracy of logarithmic dependence. Hence,
$V_e$ depends on $T_e$ linearly.

It is found out that under the mutual drag conditions, for the
drift velocities much smaller than the sound velocity in the
crystal, the $E$ dependences of $\alpha_p$ are exactly the same
for both parabolic and nonparabolic spectrum of
electrons. However, the dependences of $V_p$ are different.

Under the drag of phonons by electrons conditions, for the
scattering of electrons by DA and PA phonons, it is shown that
$\alpha_p$ and $V_p$ grow more rapidly as $E$ increases for the
nonparabolic spectrum of electrons.

In the thermal drag case, the dependences of $\alpha_p$ and
$V_p$ on $E$ are the same independent of the type of interaction
of electrons by DA or PA phonons.

In the case of drag of electrons by phonons with common drift
velocities of phonons, the dependence of $V_p$ on $E$ is the
same for both parabolic and nonparabolic spectrum of electrons,
whereas $\alpha_p$ grows faster for the nonparabolic spectrum case.

\subsection*{Acknowledgments}
This work was partially supported by the Scientific and Technical
Research Council of Turkey (TUBITAK). In the course of this work, 
T. M. Gassym was supported by a TUBITAK-NATO fellowship.

\begin{table}[h]
\renewcommand{\arraystretch}{2.4}
\addtolength{\arraycolsep}{18pt}
$$
\begin{array}{|c|c|c|}
\hline
      ~~ &  s=\frac{1}{2}  &  s=1  \\ \hline \hline
$Case i$  & \vartheta_e \sim E^{4/3} & \vartheta_e \sim E^{1/2} \\\hline
$Case ii$  & \vartheta_e \sim E^{1/3} & \vartheta_e \sim E^{1/5} \\ \hline
$Case iii$  & \vartheta_e \sim E^{4/11} & \vartheta_e \sim E^{2/9}\\ \hline
\end{array}
$$
\caption{Dependences of $\vartheta_e$ on $E$ in the condition
$\gamma_0 \rightarrow 1$.}
\renewcommand{\arraystretch}{2.4}
\addtolength{\arraycolsep}{-18pt}
\end{table}

\begin{table}[h]
\renewcommand{\arraystretch}{1.6}
\addtolength{\arraycolsep}{16pt}
$$
\begin{array}{|c|c|c|c|}
\hline
      ~ &~       & s=\frac{1}{2} & s=1   \\ \hline \hline
      ~ &V_e     & \sim ~E^{4/3} & \sim ~E^{1/2} \\
$Case i$ &\alpha_p & \sim ~E^2 & \sim ~E^{3/2} \\
      ~ &V_p     & \sim ~E^{10/3} & \sim ~E^2 \\\hline
      ~ &V_e     & \sim ~E^{1/3} & \sim ~E^{1/5} \\
$Case ii$ &\alpha_p & \sim ~E^{1/2} & \sim ~E^{3/5} \\
      ~ &V_p     & \sim ~E^{5/6} & \sim ~E^{4/5} \\\hline
      ~ &V_e     & \sim ~E^{4/11} & \sim ~E^{2/9} \\
$Case iii$ &\alpha_p & \sim ~E^{6/11} & \sim ~E^{2/3}  \\
      ~ &V_p     & \sim ~E^{10/11} & \sim ~E^{8/9} \\
\hline
\end{array}
$$
\caption{Dependences of $V_e$, $\alpha_p$ and $V_p$ on $E$ in the
condition $\gamma_0 \rightarrow 1$.}
\renewcommand{\arraystretch}{1.6}
\addtolength{\arraycolsep}{-16pt}
\end{table}

\newpage
\begin{table}[ht]
\renewcommand{\arraystretch}{2}
\addtolength{\arraycolsep}{16pt}
$$
\begin{array}{|c|c|c|c|}
\hline
$Interaction$ &     &s=\frac{1}{2} &s=1 \\ \hline \hline
$DA interaction of electrons with$ & V_e & \sim E^{4/9} & \sim E^{2/9} \\
$acoustical phonons $~ (t=1,~ r=-1/2) &~~ ~~ &~~ ~~ &~~ ~~ \\\hline
$PA interaction $~ (t=-1,~ r=1/2) & V_e & \sim E^{4/7} &
\sim E^{2/7} \\~   ~~~&~~~   ~~~&~~~   ~~~&~~~   ~~~ \\ \hline
$The momentum scattering of electrons$ & V_e & \sim E^{4/3} & \sim E^{1/2}
\\
$by impurity ions $~ (r=3/2) &~~ ~~ &~~ ~~ &~~ ~~ \\\hline
\end{array}
$$
\caption{Dependences of $V_e$ on $E$ in the condition $\gamma_0 \ll 1$.}
\renewcommand{\arraystretch}{2}
\addtolength{\arraycolsep}{-16pt}
\end{table}

\newpage
\begin{thebibliography}{99}
\bibitem{1} C. W. J. Beenakker and A. A. M. Staring, Phys. Rev. B 
{\bf 46}, 9667 (1992).
\bibitem{2} L. W. Molenkamp, A. A. M. Staring, B. W. Alphenaar and
H. van Houten, {\it Proc. 8th Int. Conf. on Hot Carriers in 
Semiconductors} (Oxford, 1993).
\bibitem{3} M. J. Kearney and P. N. Butcher, J. Phys. C {\bf 19}, 
5429 (1986); {\it ibid.} {\bf 20}, 47 (1987).
\bibitem{4} R. J. Nicholas, J. Phys. C {\bf 18}, L695 (1985).
\bibitem{5} R. Fletcher, J. C. Maan, and G. Weimann, Phys. Rev. B 
{\bf 32}, 8477 (1985). 
\bibitem{6} R. Fletcher, J. C. Maan, K. Ploog, and G. Weimann, 
Phys. Rev. B {\bf 33}, 7122 (1986).
\bibitem{7} D. G. Cantrell and P. N. Butcher, J. Phys. C {\bf 19}, 
L429 (1986); {\it ibid.} {\bf 20}, 1985 (1987); {\it ibid.}
{\bf 20}, 1993 (1987).
\bibitem{8} L. D. Hicks and M. S. Dresselhaus, Phys. Rev. B {\bf 47}, 
12727 (1993).
\bibitem{9} X. Zianni, P. N. Butcher, and M. J. Kearney, Phys. Rev.
B {\bf 49}, 7520 (1994).
\bibitem{10} R. Fletcher, J. J. Harris, C. T. Foxon, M. Tsaousidou, 
and P. N. Butcher, Phys. Rev. B {\bf 50}, 14991 (1994).
\bibitem{11} X. L. Lei, J. Phys.: Condens. Matter {\bf 6}, L305 (1994).
\bibitem{12} D. Y. Xing, M. Liu, J. M. Dong, and Z. D. Wang, Phys. 
Rev. B {\bf 51}, 2193 (1995).
\bibitem{13} X. L. Lei, J. Cai, and L. M. Xie, Phys. Rev. B 
{\bf 38}, 1529 (1988).
\bibitem{14} D. N. Zubarev, {\it Nonequilibrium Statistical 
Thermodynamics}, (New York, Consultants Bureau, 1974).
\bibitem{15} E. M. Conwell and J. Zucker, J. Appl. Phys.
{\bf 36}, 2192 (1995).
\bibitem{16} A. A. Abrikosov, {\it Introduction to the Theory of
Normal Metals: Solid State Physics Suppl} (New York,
Academic, 1972), Vol. 12.
\bibitem{17} B. M. Askerov, {\it Electron Transport Phenomena in
Semiconductors}, (Singapore, World Scientific, 1994).
\bibitem{18} M. Bailyn, Phys. Rev. {\bf 112}, 1587 (1958); 
{\it ibid}. {\bf 157}, 480 (1967).
\bibitem{19} L. E. Gurevich and T. M. Gassymov, Fizika Tverd. 
Tela {\bf 9}, 3493 (1967).
\bibitem{20} M. M. Babaev and T. M. Gassymov, Phys. Stat. 
Solidi(b) {\bf 84}, 473 (1977).
\bibitem{21} M. M. Babaev and T. M. Gassymov, Fizika Technika
Poluprovodnikov {\bf 14}, 1227 (1980).
\bibitem{22} T. M. Gassymov, A. A. Katanov and  M. M. Babaev,
Phys. Stat. Solidi(b) {\bf 119}, 391 (1983).
\bibitem{23} M. M. Babaev, T. M. Gassymov and A. A. Katanov,
Phys. Stat. Solidi(b) {\bf 125}, 421 (1984).
\bibitem{24} X. L. Lei, C. S. Ting, Phys. Rev. B {\bf 30}, 4809 
(1984); {\bf 32}, 1112 (1985).
\bibitem{25} T. H. Geballe and G. W. Hull, Phys. Rev. {\bf 94}, 
279 (1954); {\it ibid}. {\bf 94}, 283 (1954).
\bibitem{26} M. W. Wu, N. J. M. Horing and H. L. Cui,
cond-mat 9512114.
\bibitem{27} T. M. Gassymov, A. A. Katanov, J. Phys.: Condens. 
Matter {\bf 2}, 1977 (1990).
\bibitem{28} T. M. Gassymov, in the book: {\it Nekotorye Voprosy
Eksp. Teor. Fiz.}, (Baku, Elm, 1977), p. 3; Doklady Akademy Nauk
Azerbaijan SSR, Seri. Fiz. Math. Nauk {\bf 32} (6), 19 (1976). 
\bibitem{29} T. M. Gassymov, in the book: {\it Nekotorye Voprosy
Teor. Fiz.}, (Baku, Elm, 1990).
\bibitem{30} T. M. Gassymov, Doklady Akademy Nauk Azerbaijan SSR, 
Seri. Fiz. Math. Nauk {\bf 32} (6), 3 (1976); T. M. Gassymov and 
M. Y. Granowskii, Izv. Akad. Nauk Azerbaijan SSR, Seri. Fiz. Tekh. 
Math. Nauk. {\bf 1}, 55 (1976).
\bibitem{31} I. G. Kuleev, I. I. Lyapilin, A. A. Lanchakov, and 
I. M. Tsidil'kovskii, Zh. Eksp. Teor. Fiz. {\bf 106}, 1205 (1994) 
[JETP {\bf 79}, 653 (1994)]. 
\bibitem{32} I. I. Lyapilin and K. M. Bikkin, in {\it Proceedings 
of the 4th Russia Conference on Physics of Semiconductors, 
Novosibirsk, 1999}, pp. 52. 
\bibitem{33} I. I. Lyapilin and K. M. Bikkin, Fiz. Tekh. Poluprovodn. 
(St. Petersburg), {\bf 33} (6), 701 (1999) [Semiconductors {\bf 33}, 
648 (1999)].
\bibitem{34} I. G. Kuleev, A. T. Lonchakov, I. Yu. Arapova and G. I. 
Kuleev, Zh. Eksp. Teor. Fiz. {\bf 114}, 191 (1998) 
[JETP {\bf 87}, 106 (1998)].
\bibitem{35} S. S. Shalyt and, S. A. Aliev, Fiz. Tverd. Tela 
{\bf 6} (7), 1979 (1964).
\bibitem{36} S. A. Aliev, L. L. Korenblit, and S. S. Shalyt, 
Fiz. Tverd. Tela {\bf 7} (6), 1973 (1965).
\bibitem{37} I. N. Dubrovnaya and Yu. I. Ravich, Fiz. Tverd. 
Tela {\bf 8} (5), 1455 (1966).
\bibitem{38} V. I. Tamarchenko, Yu. I. Ravich, L. Ya Morgovskii 
{\it et al.}, Fiz. Tverd. Tela {\bf 11} (11), 3506 (1969).
\bibitem{39} K. M. Bikkin, A. T. Lonchakov, and I. I. Lyapilin, 
Fiz. Tverd. Tela {\bf 42} (2), 202 (2000) [Phys. Solid State, 
{\bf 42} (2), 207 (2000)].
\bibitem{40} L. E. Gurevich and T. M. Gassymov, Fiz. Tverd. Tela 
{\bf 9}, 106 (1967).
\end{thebibliography}
\end{document}